\documentclass[12pt]{article}
\usepackage[dvips]{color}
\usepackage{epsfig}
\usepackage{amsmath}
\usepackage{graphicx}

\begin{document}
\title{The lower bound violation of shear viscosity to entropy ratio due to logarithmic correction in STU model}
\author{{Behnam Pourhassan$^{a}$\thanks{Email: b.pourhassan@du.ac.ir} and Mir Faizal$^{b,c,}$\thanks{Email: f2mir@uwaterloo.ca}}\\
$^{a,}${\small {\em School of Physics, Damghan University, Damghan, Iran}}\\
$^{b,}${\small {\em  Irving K. Barber School of Arts and Sciences,  University of British Columbia - Okanagan,}}\\
{\small {\em Kelowna,   BC V1V 1V7, Canada}}
\\
$^{c,}${\small {\em  Department of Physics and Astronomy, University of Lethbridge,}}\\
{\small {\em Lethbridge, AB T1K 3M4, Canada}}
}
\date{}
\maketitle

\begin{abstract}
In this paper, we analyze the effects of thermal fluctuations on a STU black hole.
We observe that these thermal fluctuations can affect the stability of a STU black hole.
We will also analyze the effects of these thermal fluctuations on the thermodynamic of STU black hole.
Furthermore, in the Jacobson formalism such a modification will produce a deformation of the geometry of the STU black hole. Hence, we use the
AdS/CFT correspondence to analyze the effect of such a deformation on the dual quark-gluon plasma. So,
we explicitly analyze the effect of thermal fluctuations  on the shear
viscosity to entropy ratio in the quark-gluon plasma,
and analyze the effects of thermal fluctuations on this ratio.
\\\\
{\bf Keywords:}  AdS/CFT; Quantum Gravity; Thermodynamics.\\\\
{\bf Pacs Number:}  04.60.-m 04.70.-s 05.70.-a
\end{abstract}

\section{Introduction}

The AdS/CFT correspondence relates the supergravity solution in the AdS space to the conformal field theory (CFT)
on its boundary \cite{mald}. As the AdS/CFT correspondence relates the AdS geometry to the ground
state of a conformal field theory, a deformation of the AdS solution in the bulk will also modify the CFT dual to that
AdS solution. In fact, such a deformation will result in the excitation of the ground state of the dual CFT solution.
So, a black hole in AdS space corresponds to heating up the system, and this in turn corresponds to exciting the ground state
of the CFT.
In this paper, we analyze an interesting  non-extremal black hole solutions which is motivated from results obtained using
the string theory, and it called the STU black hole solution \cite{9st,10st}.  The STU black holes solution  can be considered as the holographic dual of quark-gluon plasma (QGP), and it is possible to study QGP using STU/CFT correspondence \cite{15st}.  The QGP is a phase in quantum chromodynamics (QCD) which exists at extremely high temperature and  density. There are many important quantities in QGP such as the  shear viscosity,
drag force and jet-quenching and they can be calculated holographically from a STU black hole \cite{11st,   13st, 14st}.
So, in this paper, we will first analyze a deformation of the STU black hole geometry.
Then, we will analyze the modification to the QGP because of such a deformation of the STU geometry using the STU/CFT correspondence.
Specially, we correct the shear viscosity to entropy ratio. It is conjectures that mentioned ratio has a universal value $\frac{1}{4\pi}$ in natural units. In fact it suggests a lower bound so we have,

$$\frac{\eta}{s}\geq\frac{1}{4\pi}.$$

However, due to various effects this lower bound may be violated \cite{11,22,33,44,55,66,77}.

In order to analyze the deformation of the STU black hole geometry by thermal fluctuations, we need to first understand the relation between
geometry of a black hole and its thermodynamics. In that case, the thermodynamics of STU black holes have been studied originally by the Refs. \cite{95, 70}. The area-entropy relation establishes a relation between the geometry of space-time and thermodynamics
of a black hole
 \cite{1, 1a}. According to the area-entropy relation
the  entropy   of  a black holes scales with the area of its horizon   \cite{2, 4, 4a}. It may be noted that this  observation has led to the development of the  holographic principle \cite{5, 5a}, and AdS/CFT correspondence (which has motivated the STU/CFT or STU/QCD correspondence \cite{11st,   13st, 14st, 15st}) is based on the holographic
principle. This is because the holographic principle related the degrees of freedom in any region of space to the degrees of freedom
on the boundary surrounding that region of space. This relation between the geometry of space-time and thermodynamics is more evident in the Jacobson formalism where Einstein equation is viewed as a  thermodynamics relation
\cite{z12j, jz12}. In fact, the Einstein's equation is derived in the Jacobson formalism  by requiring
Clausius relation to hold for all the local Rindler causal horizons through each space-time point.
As the  Jacobson formalism establishes a clear relation between the geometry of space-time and thermodynamics,
quantum fluctuations in the geometry of space-time will produce thermal fluctuations in the thermodynamics
of black holes in the Jacobson formalism. Thus, we expect   the thermodynamics of all black holes
to get corrected because of the thermal fluctuations in the Jacobson formalism.

It has been demonstrated that the area-entropy relation gets modified due to thermal fluctuations  \cite{l1, SPR}.
These corrections to the area-entropy relation have been studied using both analyzing the fluctuations in the  energy
of the system, and relating this system to a conformal field theory. However, the quantum fluctuations become important
when the geometry is probed at very small scales, and the thermal fluctuations also become important when the temperature of
the black hole is very large, and this corresponds to a very small size of the black hole. So, when the black hole reduced in size
due to the Hawking radiation, the effects of thermal fluctuations cannot be neglected. So, when the
size of the black hole becomes of the order of Planck scale,  the temperature of the black hole become very large,
and the contribution from thermal fluctuations also becomes very important for such black holes. The effect of thermal fluctuation on a black hole in an   anti-de Sitter space-time  have been studied, and the correction to thermodynamics of such a black hole has also been obtained \cite{adbc}.
The corrected thermodynamics of such a black hole has been used for analyzing   the phase transition in that system.
The corrections to the thermodynamics of   a black Saturn  have also been studied, and it was observed that the entropy
of both the black hole and the black ring gets corrected due to thermal fluctuations \cite{dabc}. The black Saturn are thermodynamically
stable because of the rotation of the black ring. However, it is possible for charged dilatonic black Saturn to remain stable
because of background fields, and the thermodynamics of
a charged dilatonic black Saturn has  been discussed \cite{1407.2009}. The corrections to the
 thermodynamics of such a charged dilatonic black Saturn have also been analyzed using the relation between this system
 and a conformal field theory \cite{ab}. The corrections to the thermodynamics of a modified Hayward black hole have also
 been discussed, and it has been  demonstrated that the modified Hayward black hole
is stable even after the thermal fluctuations are taken into account,
as long as the event horizon is larger than a certain critical value \cite{abab}.
It has been demonstrated that for all these systems the
correction due to thermal fluctuations is  a logarithmic correction.
It may be noted that such correction terms have also been obtained from
non-perturbative quantum  general relativity \cite{1z}, Cardy formula \cite{card}, exact partition function for a  BTZ black hole \cite{gks},
and matter fields in backgrounds of a black hole  \cite{other, other0, other1}.
In fact, even corrections obtained from string theory  are   logarithmic corrections  \cite{solo1, solo2, solo4, solo5, jy, SPR, PF, PSFB}. All above studies indicated that logarithmic corrected thermodynamics of black objects is important field of study in theoretical physics.
So, in this paper, we will analyze the effect of such logarithmic corrections for a STU black hole.
The logarithmic corrections to the thermodynamics of STU black hole will deform the geometry of the STU black hole in the Jacobson formalism,
this  will directly affect the properties of QGP inspired by the AdS/CFT correspondence. So, in this paper, we will analyze the effect of such a deformation on the shear viscosity to entropy ratio of QGP. This paper is organized as follows. In the next section we review some important properties of STU black hole from thermodynamical point of view. In section 3 we introduce logarithmic correction, and in section 4 study its effect on the shear viscosity to entropy ratio. Finally in section 5 we give conclusion.

\section{STU black hole}
\begin{figure}[h!]
 \begin{center}$
 \begin{array}{cccc}
\includegraphics[width=70 mm]{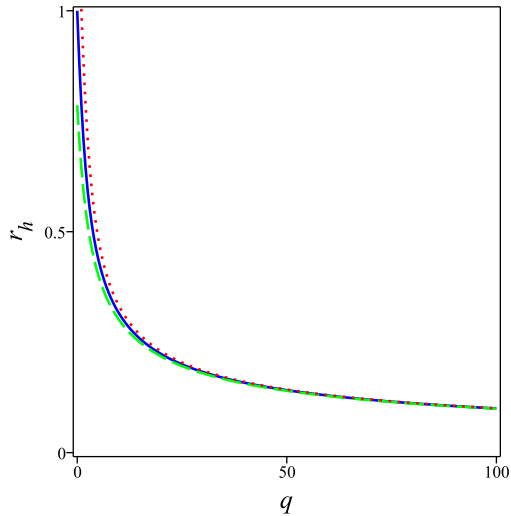}
 \end{array}$
 \end{center}
\caption{Black hole event horizon in terms of the black hole charge with $R=\mu=1$ and $k=1$ (dashed line), $k=0$ (solid line), and $k=-1$ (dotted line).}
 \label{fig1}
\end{figure}

In this section we recall STU model in five dimension including electric charge and write thermodynamical properties which is useful in the context of AdS/CFT correspondence.
The metric for the $5D$ STU model with three electrical charges can be written as
\begin{equation}\label{s1}
ds^{2}=-\frac{f_{k}}{{\mathcal{H}}^{\frac{2}{3}}}dt^{2}
+{\mathcal{H}}^{\frac{1}{3}}(\frac{dr^{2}}{f_{k}}+\frac{r^{2}}{R^{2}}d\Omega_{3,k}^{2}),
\end{equation}
where,
\begin{eqnarray}\label{s2}
f_{k}&=&k-\frac{\mu}{r^{2}}+\frac{r^{2}}{R^{2}}{\mathcal{H}},\nonumber\\
{\mathcal{H}}&=&\prod_{i=1}^{3} H_{i},\nonumber\\
H_{i}&=&1+\frac{q_{i}}{r^{2}}, \hspace{10mm} i=1, 2, 3,.
\end{eqnarray}
Here, $R$ is the radius of the AdS space and it is related  to the coupling
constant as $R=1/g$. The  coupling constant is also  related to the
cosmological constant as $\Lambda=-6g^2$. Furthermore, $r$ is the radial
coordinate of the black hole, and the  three electrical charges of black hole, corresponding to the three scalar field $X^{i}={\mathcal{H}}^{\frac{1}{3}}/H_{i}$,
are denoted by $q_{i}$, with $X^{1}X^{2}X^{3}=1$ $(STU=1)$. The non-extremality  parameter is denoted by $\mu$. The
  closed universe has    $k=1$, the flat universe has  $k=0$, and the open universe has  $k=-1$. The  metric for these universes can be written as
\begin{eqnarray}\label{s3}
d\Omega_{3,k}^{2}\equiv\big\{\begin{array}{ccc}
R^{2}(d\rho^{2}+\sin^{2}\rho d\theta^{2}+\sin^{2}\rho\sin^{2}\theta d\phi^{2}) \\
dx^{2}+dy^{2}+dz^{2} \\
R^{2}(d\rho^{2}+\sinh^{2}\rho d\theta^{2}+\sinh^{2}\rho\sin^{2}\theta d\phi^{2}) \\
\end{array}
\end{eqnarray}
for $k=1,0,-1$ respectively.
We will only consider the case where there is only one-charge for the  black hole ($q_{1}=q, q_{2}=q_{3}=0$),
and where  ${\mathcal{H}}=H=1+\frac{q}{r^{2}}$. So, the
 only free parameter of the model will be $q$. Now, the temperature and the entropy of this black hole can be written as \cite{0601144},
\begin{equation}\label{s4}
T=\frac{r_{h}}{2\pi
R^{2}}\frac{2+\frac{q+kR^{2}}{r_{h}^{2}}}{{\sqrt{1+\frac{q}{r_{h}^{2}}}}},
\end{equation}
and
\begin{equation}\label{s5}
s=\frac{r_{h}^{3}\sqrt{1-\frac{q}{r_{h}^{2}}}}{4GR^{3}},
\end{equation}
where $G$ is Newton's constant and it is related to the AdS curvature as
$G=\frac{\pi R^{3}}{2N^{2}}$. Here, $N$ is the number of colors. We should note that there is a coefficient $k^{2}$ in denominator of the equation (\ref{s5}) for the cases of open and closed universes. Hence, the equation (\ref{s5}) is valid in its present form for all cases of $k=0, \pm1$.\\
It may be noted that
$r_{h}$  is given by the root of $f_{k}=0$,
\begin{equation}\label{s6}
r_{h}=\frac{1}{2}\sqrt{2\sqrt{k^{2}R^{4}+2kqR^{2}+q^{2}+4\mu R^{2}}-2kR^{2}-2q}.
\end{equation}
The black hole horizon is a decreasing function of the black hole charge as illustrated by the Fig. {\ref{fig1}}
for different $k$. The size of the black hole will be small for large electrical charge. However,  for a  large value of the black hole
charge there is no way to distinguish between event horizons of a black hole in open, closed and flat universe.

\begin{figure}[h!]
 \begin{center}$
 \begin{array}{cccc}
\includegraphics[width=70 mm]{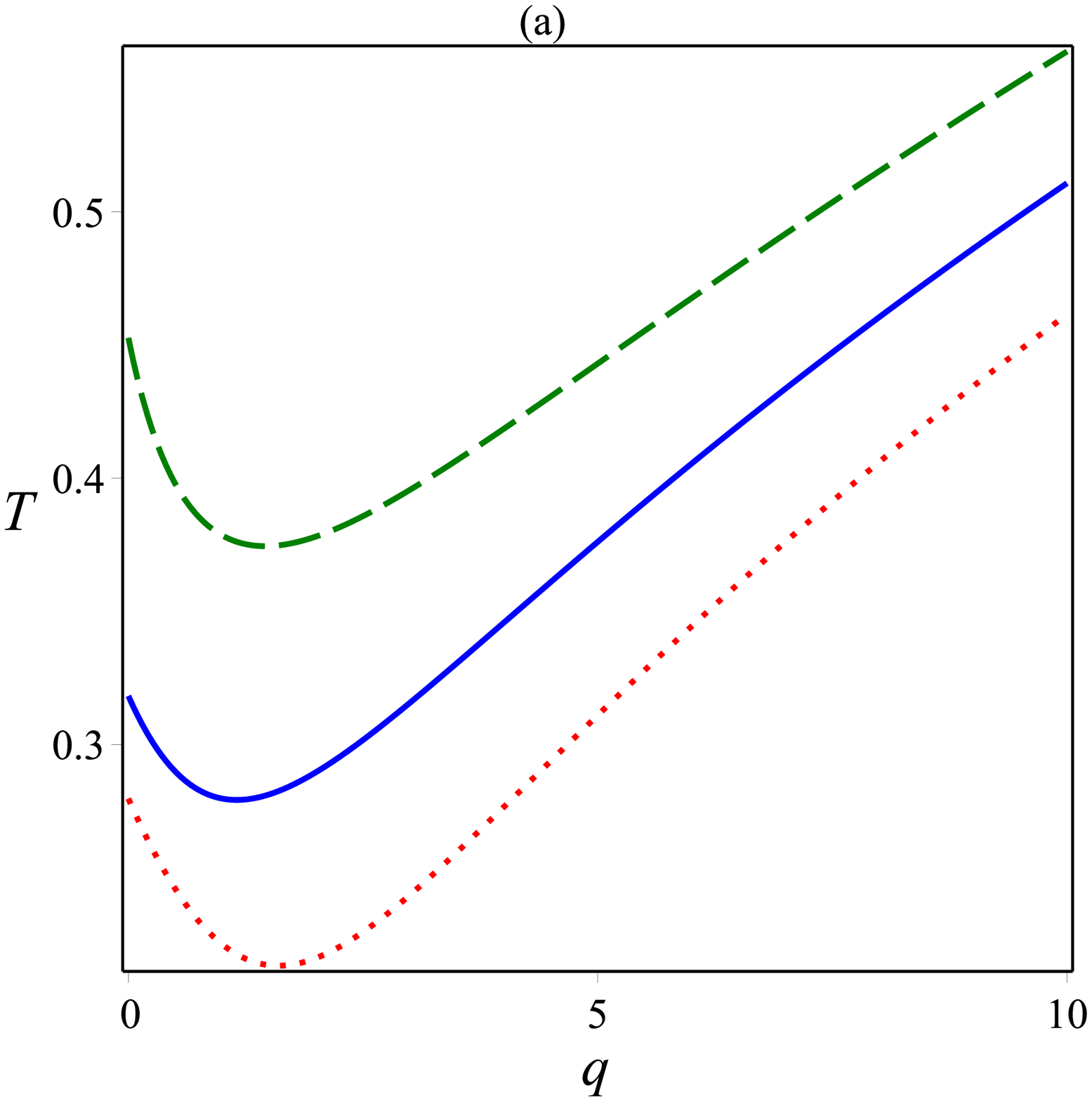}\includegraphics[width=70 mm]{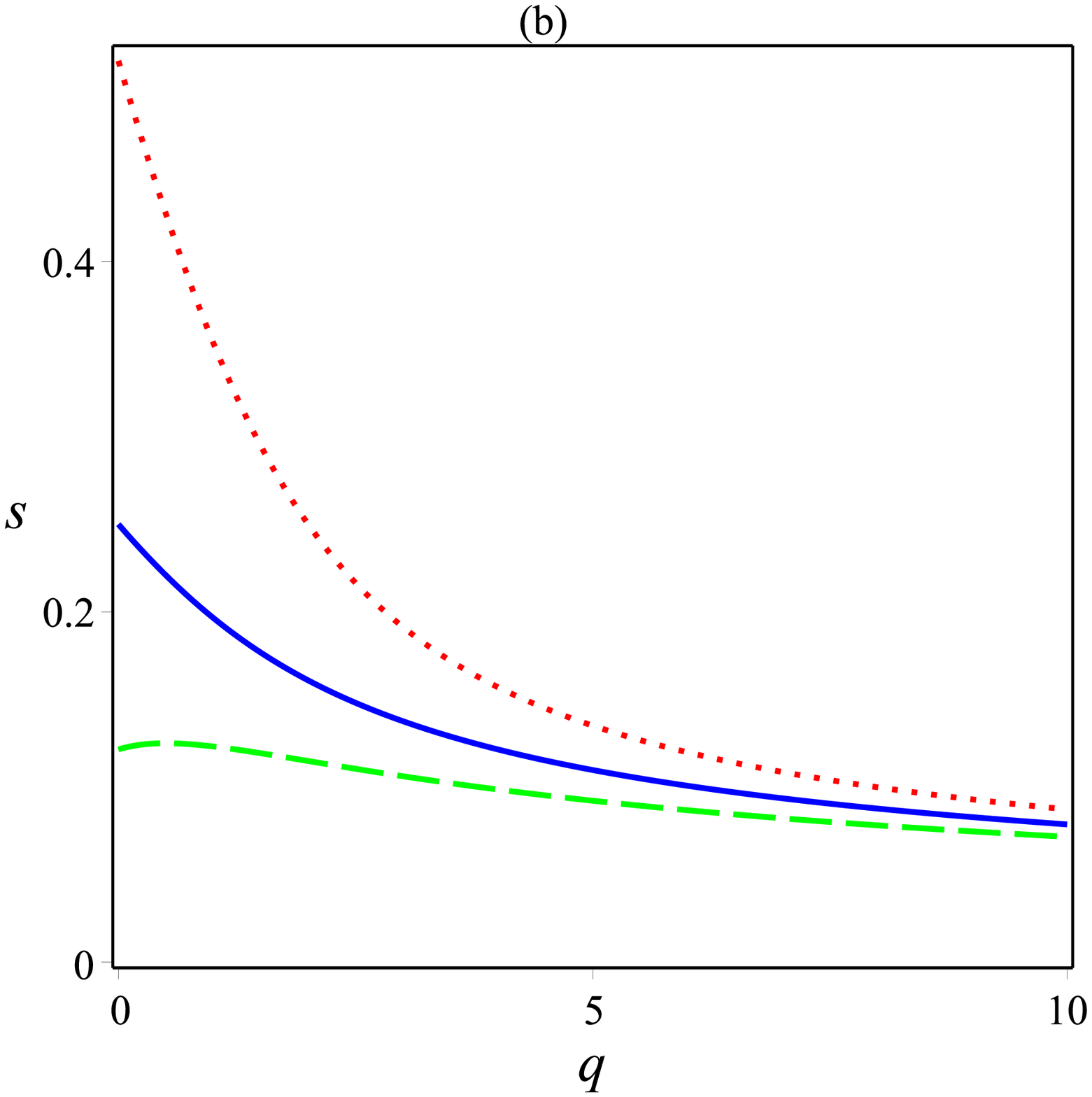}
 \end{array}$
 \end{center}
\caption{Black hole temperature (a) and entropy (b) in terms of the black hole charge with $R=G=\mu=1$ and $k=1$ (dashed line), $k=0$ (solid line),
and $k=-1$ (dotted line).}
 \label{fig2}
\end{figure}

Using the Eqs. (\ref{s4}), (\ref{s5}) and (\ref{s6}), the  temperature and entropy of the black hole can be expressed in terms of the black hole charge.
In the Fig. \ref{fig2} (a), we demonstrate  that the temperature is a decreasing function of charge for small values of $q$.
There is a critical $q_{c}$, where the black hole has a  minimum temperature. Then for $q>q_{c}$ black hole temperature increases with $q$.\\
In the Fig. \ref{fig2} (b), we observe  that by increasing the charge of the black hole, its    entropy decreases. This is also
  expected from Fig. {\ref{fig1}}, because entropy of the  black hole  is proportional to radius of the event horizon $r_{h}$.

\begin{figure}[h!]
 \begin{center}$
 \begin{array}{cccc}
\includegraphics[width=70 mm]{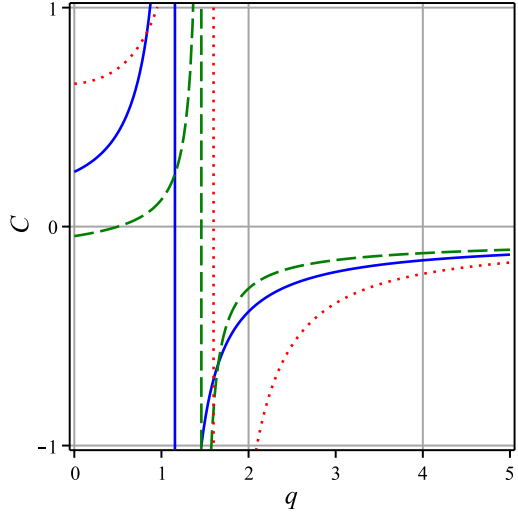}
 \end{array}$
 \end{center}
\caption{Specific heat in terms of the black hole charge with $R=G=\mu=1$ and $k=1$ (dashed line), $k=0$ (solid line), and $k=-1$ (dotted line).}
 \label{fig3}
\end{figure}

The specific heat can be written as,
\begin{equation}\label{s7}
c=T\left(\frac{\partial T}{\partial q}\right)^{-1}\left(\frac{\partial s}{\partial q}\right).
\end{equation}
In the Fig. \ref{fig3}, we plot the  specific heat
and observe  that for the large value of the black hole charge, there is an instability.
Thus, by increasing the charge of the black hole, the black hole  become smaller and warmer and it  enters an unstable phase.
However, we do not  expect such instabilities and we can remove them. The  thermal fluctuations are important when the size of black hole is small.
In the next section, we will analyze the effects of the logarithmic correction due to the thermal fluctuation,
on the specific heat.  We will observe that we can remove such instabilities by using such corrections.

We can use the following expression for the shear viscosity $\eta$ \cite{0601144},
\begin{equation}\label{s5-1}
\eta=\frac{r_{h}^{3}\sqrt{1+\frac{q}{r_{h}^{2}}}}{16\pi GR^{3}},
\end{equation}
to investigate famous ratio $\frac{\eta}{S}$, where $S$ is corrected entropy which will define in the next section.
Thus, the conjectured universal relation $\frac{\eta}{s}=\frac{1}{4\pi}$ hold for the STU model. So, using the Eqs. (\ref{s5}) and (\ref{s5-1}),  we can verify the  universal behavior. However there are some examples \cite{violate1, violate2} where mentioned ratio deviates below $\frac{1}{4\pi}$. It should be noted that the calculation of $\frac{\eta}{s}$ in the STU background was first performed by the Ref. \cite{0601144, 0601157}.

\section{Thermal fluctuations}
It is possible to analyze the effects of  thermal fluctuations  on the  black objects thermodynamics \cite{l1}.
The entropy of any black objects gets corrected by a logarithmic term due to these thermal fluctuations.
Thus, if we assume $\beta_{\kappa}^{-1} = T$ as a temperature close to the
equilibrium, and $\beta_0^{-1} = T_0 $ as the equilibrium temperature, then  the corrected entropy can be written as \cite{SPR}
\begin{equation}\label{s9}
 S = s - \frac{\ln  s^{\prime\prime}}{2},
\end{equation}
where
\begin{equation}\label{s10}
s^{\prime\prime}=  ({\partial^2 s}/{\partial \beta_{\kappa}^2} )|_{\beta_{\kappa} = \beta_0}.
\end{equation}
It is also possible to  express the  second derivative of the entropy  in terms of the fluctuations of the energy  near the equilibrium.
Thus, the     corrected  entropy can be written as \cite{l1, dabc, Prad}
\begin{equation}\label{s11}
S = s -\frac{\alpha}{2} \ln |cT^{2}|+\cdots,
\end{equation}
where $s$ is  original entropy and and $c$ is the original specific heat of the system.
Furthermore, as almost all approaches to quantum gravity generate such a    logarithmic correction,
however, the coefficient of such a correction term depends on the exact model of quantum gravity that has been used.
Thus, such a coefficient can be used as a parameter than can test different model of quantum gravity.
This is because  different approaches to quantum gravity would generate different values of  the
 coefficient of the logarithmic term. So, in this paper, we will
 keep this analysis general and introduce a general parameter
$\alpha$, which will be the coefficient of the logarithmic correction term.
Now when $\alpha = 1$, the usual  thermal fluctuations taken into account, which is corresponding to a very small black object.
On the other hand for $\alpha =0$, thermal fluctuations ignored, which is corresponding to
the large black objects. Finally dots denotes higher order corrections which may be considered in future works.

It is also possible to relate the microscopic degrees of freedom of a black hole with a conformal field theory \cite{ab}. Thus, using the modular invariance of the partition function of the conformal field theory, corrected  entropy can be written as \cite{l1, ab}
\begin{equation}\label{s11-1}
S = s -\frac{\alpha}{2} \ln |sT^{2}|+\cdots,
\end{equation}
It may be noted that  for the charged STU, there is important difference between results obtained from relations (\ref{s11}) and (\ref{s11-1}),
however both are the same at $q=0$. So, there is a difference between the corrections generated from a conformal field theory, and
the corrections generated from the fluctuations in the energy of the system.
So, we observe that the effect of thermal fluctuations for the STU black holes is
differently from the effect of thermal fluctuation on most other black hole solutions.  This is because
correction from both these approaches generated the same effects for all the other black hole that have been analyzed using this formalism
\cite{l1,SPR, adbc, dabc, ab, abab}.

\begin{figure}[h!]
 \begin{center}$
 \begin{array}{cccc}
\includegraphics[width=70 mm]{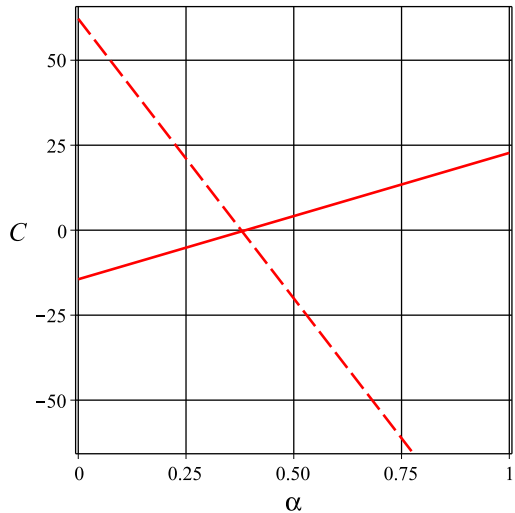}
 \end{array}$
 \end{center}
\caption{Corrected specific heat in terms of $\alpha$ with $R=G=\mu=1$ and $k=0$. Dashed line denotes the case of $q<q_{c}$ and solid line denotes the case of $q\geq q_{c}$.}
 \label{fig4}
\end{figure}

Now using the logarithmic corrected entropy (\ref{s11}),  we can obtain the corrected specific heat as
\begin{equation}\label{s12}
C=T\left(\frac{\partial T}{\partial q}\right)^{-1}\left(\frac{\partial S}{\partial q}\right).
\end{equation}
In the Fig. \ref{fig4}, we can observe the behavior of corrected specific heat for one-charged STU black hole,
in terms of $\alpha$ for flat universes. For the reason we explain later, we only consider the case of $k=0$. However, we have the same situation for the open and closed universes. As we see already, some instability happened for the large black hole charge. We show that there is a critical temperature $q_{c}$ so for the case of $q>q_{c}$ the black hole is unstable, while in presence of logarithmic correction ($\alpha\neq0$) with appropriate choice of $\alpha_{c}$ the black hole is stable. For the selected value of parameters ($R=G=\mu=1$) we can see that $q_{c}\approx1.16$ and $\alpha_{c}\approx 0.39$. Solid line of the Fig. \ref{fig4} show black hole specific heat for large electric charge. It is clear that the black hole is unstable for $\alpha=0$ as mentioned already by the Fig. \ref{fig3}. We can see that for the appropriate choice such as $\alpha=1$ we have totally stable black hole. Therefore we find that the logarithmic correction help to gain stability of black hole at high temperature.

\section{Shear viscosity to entropy ratio}
In this section, we are going to study the effect of the logarithmic correction on
the shear viscosity to entropy ratio. We will consider three different cases corresponding to thermal fluctuation effects.\\

\begin{figure}[h!]
 \begin{center}$
 \begin{array}{cccc}
\includegraphics[width=50 mm]{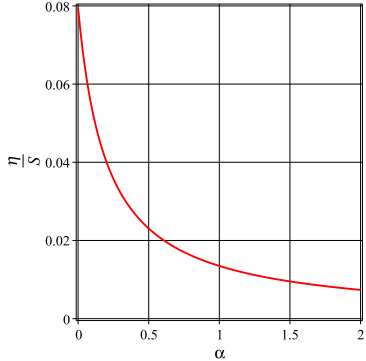}
 \end{array}$
 \end{center}
\caption{Shear viscosity to entropy ratio in terms of $\alpha$ with $R=G=\mu=1$ and $k=0$.}
 \label{fig5}
\end{figure}

So,  first of all we can make a simple assumption, i.e.,
we can assume that the thermal fluctuations  do not affect
the shear viscosity,
hence, we can obtain  shear viscosity to entropy ratio using corrected entropy.
So,  we can use corrected entropy given by the Eq. (\ref{s11}) and the Eq.
(\ref{s5-1}), to obtain  the corrected shear viscosity to entropy ratio.
In the case of $\alpha=0$, we have $\frac{\eta}{s}=\frac{1}{4\pi}$. However,
in  presence of logarithmic correction, we obtain
\begin{equation}\label{main}
\frac{\eta}{S}=\frac{r_{h}^{3}\sqrt{1+\frac{q}{r_{h}^{2}}}}{4\pi r_{h}^{3}\sqrt{1+\frac{q}{r_{h}^{2}}}-8\pi GR^{3}\alpha\ln{cT^{2}}}.
\end{equation}
So,  from   the Fig. \ref{fig5},  we can observe the effect of $\alpha$ on the shear viscosity to entropy ratio. Thus, we observe that lower bound ($\frac{1}{4\pi}=0.08$) decreased due to the logarithmic correction.
It may be noted that using  the AdS/CFT correspondence, these  corrections in the bulk correspond to $1/N^{2}$ corrections in the dual boundary theory. It is clear that $\alpha=0$ yields to conjectured universal lower bound ($\frac{\eta}{s}=\frac{1}{4\pi}=0.08$) while $\alpha=1$ yields to $\frac{\eta}{s}=0.01-0.03$, which means universal lower bound violated. It should be noted that shear viscosity is typically defined in flat space ($k=0$) hence in this section we only consider the case of $k=0$. As we mentioned, the logarithmic correction is correspond to $1/N^{2}$ correction, hence $G R^{3}\alpha$ should be a small number proportional to $1/N^{2}$ and small-alpha region of the Fig. \ref{fig5} is in any way reliable. Hence, we can rewrite the equation (\ref{main}) as follow,
\begin{equation}\label{main-2}
\frac{\eta}{S}=\frac{1}{4\pi}+\frac{\gamma}{N^{2}}\frac{\ln{cT^{2}}}{r_{h}^{2}\sqrt{q+r_{h}^{2}}},
\end{equation}
where $\gamma$ is small positive constant.
It is clear that the shear viscosity to entropy ratio is decreasing function of $\gamma$,
hence lower bound is violated.\\

It is possible to obtain another result by using
 a better and more physical approximation.  In such a calculation
the thermal fluctuations  also correct the shear viscosity. In fact, as the corrections
to the entropy are $\frac{1}{N^{2}}$ correction in the bulk, we expect such
corrections to also correct the shear viscosity. It is possible to suggest the
corrected value of shear viscosity, and this corrected value of the shear viscosity can correct the
shear viscosity to entropy ratio.
In that case, shear viscosity to entropy ratio may given by,
\begin{equation}\label{main-4}
\frac{\eta}{S}=\frac{\frac{r_{h}^{3}}{4GR^{3}}\sqrt{1+\frac{q}{r_{h}^{2}}}+\mathcal{O}(\alpha)}{4\pi \left(\frac{r_{h}^{3}}{4GR^{3}}\sqrt{1+\frac{q}{r_{h}^{2}}}-\frac{\alpha}{2}\ln{cT^{2}}\right)}.
\end{equation}
It is clear that the above ratio may increasing or decreasing
function of $\alpha$ and also may be constant for suitable choice of $\mathcal{O}(\alpha)$, hence the lower bound for this ratio may hold.
However, for the appropriate value of $\mathcal{O}(\alpha)$, lower
bound may violated and the shear viscosity to entropy ratio yields to zero.\\
The best way to calculate $\mathcal{O}(\alpha)$ is Kubo formula which relates the shear viscosity to the correlation function
of the stress-energy tensor at zero spatial momentum by using the retarded Green's function \cite{0601157}.\\

It is also possible to obtain an expression for the
corrected  shear viscosity, such that the ratio of the correct viscosity and
corrected entropy is still does not violate conjectured
universal minimum bound. This can be used to understand the behavior
of shear viscosity in this limiting case.
Thus,  we can
assume that universal value $\frac{\eta}{S}=1/4\pi$ hold for the corrected case,  and obtain
\begin{equation}\label{main-5}
\frac{\eta}{S}=\frac{\eta}{\frac{r_{h}^{3}\sqrt{1+\frac{q}{r_{h}^{2}}}}{4GR^{3}}-\frac{\alpha}{2}\ln{cT^{2}}}=\frac{1}{4\pi}.
\end{equation}
Therefore, we can obtain corrected shear viscosity as,
\begin{equation}\label{main-6}
\eta=\frac{r_{h}^{3}\sqrt{1+\frac{q}{r_{h}^{2}}}}{16\pi GR^{3}}-\frac{1}{8\pi} \alpha\ln{cT^{2}}.
\end{equation}
It may be noted that, like all gauge theories with Einstein gravity dual, a lower bound hold for all values of $\alpha$. However, by study perfect quark-gluon liquid \cite{25}, it has been found enhanced viscosity to entropy ratio $\frac{5}{8\pi}$. It has been also found that higher curvature corrections in the dual gravitational theory modify this ratio \cite{66}, hence higher derivative corrected STU black hole \cite{G} is interesting issue to investigate under logarithmic correction. It has been also argued that for certain corrected theories the lower bound violated. Just like the logarithmic corrected case we show that the lower bound may violated due to thermal fluctuations.

\section{Conclusion}
In this paper, we have analyzed a special case of STU black hole in five dimensions with an electric charge. We have used AdS/CFT to investigate the effect of thermal fluctuations on the properties of QGP, specially the shear viscosity to entropy ratio.
First of all we used logarithmic corrected entropy and studied thermodynamics of one-charged STU black hole.
We found that the logarithmic corrections affect the black hole stability. For instance, instability of the black hole at high temperature changes to stable phase in presence of logarithmic correction.
We demonstrated  that the viscosity to entropy ratio of QGP dual of STU background is
reduced due to thermal fluctuations with appropriate choice of correction parameter. Our study was based on three different assumptions. First, we assumed that the shear viscosity not changed due to logarithmic correction and only used logarithmic correction of the entropy and found that positive value of $\alpha$ yields to violation of lower bound. In the second way we write general form of the corrected shear viscosity and claim that appropriate value of $\alpha$ gives the lower bound violation. Finally we assumed universal value and calculated corrected shear viscosity.
The corrections to the thermodynamics can be obtained by analyzing the fluctuations in the energy of the system.
They can also be analyzed using the relation of the black hole microstates with a conformal field theory. It has been
observed that the effects of thermal fluctuations from both of these approaches are the same for all black hole solutions that have
been previously analyzed using this formalism \cite{l1,SPR, adbc, dabc, ab, abab}.
However, in this paper, it was observed that the effects of thermal fluctuation for a charged STU black hole from the energy fluctuations
is different from the effect of thermal fluctuations for a STU black hole obtained from a conformal field theory. However, both of these
effect are the same when the charge vanishes. It might be interesting to investigate the reason for this further.

It may be noted that U-duality invariant expression for the area-entropy relation has been obtained for a  stationary, asymptotically flat, non-extremal STU black holes \cite{no}. It was demonstrated that this expression can be expressed in  terms of asymptotic charges of this stationary, asymptotically flat, non-extremal STU black holes. This involves the scalar charges of the black hole which can be solved in terms of the dyonic charges and the mass of the black hole.
It might be also possible to express the corrections to the area-entropy relation in such a  stationary, asymptotically flat, non-extremal STU black holes using
 asymptotic charges. It would thus be interesting to analyze thermal fluctuations of such a STU black hole and discuss these corrections to the entropy using asymptotic charges.  Testing quantum gravity using black objects \cite{Annals} like STU black hole is also interesting research field.

The extended thermodynamics of a STU black hole has also been studied \cite{on}. This was done by viewing the cosmological constant as a thermodynamic variable of STU black hole. A fixed charge ensemble was used to perform this analysis. It was demonstrated that the phase structure associated with this black hole was conjectured to a dual  RG-flow on the space of field theories. It was also observed that the phase structure of this system resembles a Van der Waals gas for certain charge configurations. Thus, for this  system  a family of the first order phase transitions existed. Furthermore, at a critical temperature, these first order phase transitions ended  in a second order phase transition. The holographic entanglement entropy for such  charge configurations was also obtained.
    It was observed that this entanglement entropy also predicted  a transition at the critical temperature. So, the entanglement entropy can be used for analyzing the  system in an extended phase structure. In this analysis,
     holographic heat engine dual to STU black holes were also studied.
    It would be interesting to analyze the effect of thermal fluctuations on such a system.
    Thus, we could analyze the effects of thermal fluctuation on both Van der Waals gas, and the holographic entanglement entropy
    of black holes.  It would also be interesting to analyze the effect of thermal fluctuation on the
    holographic heat engines dual to a STU black holes.

\end{document}